\documentstyle[12pt,aaspp4]{article}

\def\1115{PG~1115$+$080}
\newcommand\al{\alpha}
\newcommand\ep{\epsilon}
\newcommand\th{\theta}
\newcommand\te{\th_\ep}

\newcommand\Hunits{km\ s$^{-1}$\ Mpc$^{-1}$}
\newcommand\kpc{{\rm kpc}}
\newcommand\kgal{\kappa_{gal}}

\newcommand\kgrp{\kappa_{grp}}
\newcommand\ggrp{\gamma_{grp}}
\newcommand\dgrp{d_{grp}}
\newcommand\bgrp{b_{grp}}
\newcommand\tgrp{\th_{grp}}
\newcommand\chiN{\chi^2/N_{dof}}
\newcommand\dtBC{(\Delta\tau_{BC}/25.0\ {\rm days})}
\newcommand\RABC{r_{ABC}}
\newcommand\h{h_{BC}}
\newcommand\s{\phantom{-}}
\def\refeq#1{eq.\ (\ref{eq:#1})}
\def\reffig#1{Figure \ref{fig:#1}}
\def\reffigs#1#2{Figures \ref{fig:#1} and \ref{fig:#2}}
\def\reftab#1{Table \ref{tab:#1}}

\def\deg{\ifmmode ^{\circ}
	\else $^{\circ}$\fi}
\newcommand\pdeg{\ifmmode $\setbox0=\hbox{$^{\circ}$}\rlap{\hskip.11\wd0 .}$^{\circ}
	\else \setbox0=\hbox{$^{\circ}$}\rlap{\hskip.11\wd0 .}$^{\circ}$\fi}
\newcommand\arcs{\ifmmode {^{\scriptscriptstyle\prime\prime}}
	\else $^{\scriptscriptstyle\prime\prime}$\fi}
\newcommand\arcm{\ifmmode {^{\scriptscriptstyle\prime}}
	\else $^{\scriptscriptstyle\prime}$\fi}
\newdimen\sa  \newdimen\sb
\newcommand\parcs{\sa=.07em \sb=.03em
	\ifmmode $\rlap{.}$^{\scriptscriptstyle\prime\kern -\sb\prime}$\kern -\sa$
	\else \rlap{.}$^{\scriptscriptstyle\prime\kern -\sb\prime}$\kern -\sa\fi}

\begin{document}

\title{Determining the Hubble Constant from the \\
			 Gravitational Lens \1115}
\author{C.R.~Keeton}
\author{C.S.~Kochanek}
\affil{Harvard-Smithsonian Center for Astrophysics, MS-51\protect \\ 
			 60 Garden Street \protect \\
			 Cambridge, MA 02138 }
\authoremail{ckeeton@cfa.harvard.edu}

\begin{abstract}
For the quadruple gravitational lens \1115, we combine recent
measurements of the time delays with new lens models to determine
the Hubble constant $H_0$.  We explore the effects of systematic
uncertainties in the lens models on the estimates of $H_0$, and
we discuss how the uncertainties can be reduced by future
observations.  We find that the lens cannot be
fit by an isolated lens galaxy, but that it can be well fit by
including a perturbation from the nearby group of galaxies.  To
understand the full range of systematic uncertainties it is
crucial to use an ellipsoidal galaxy and to let the group
position vary.  In this case, the existing constraints cannot
break degeneracies in the models with respect to the profiles
of the galaxy and group and to the position of the group.
Combining the known time delays with a range of lens models
incorporating most of the plausible systematic effects
yields $H_0=51_{-13}^{+14}$ \Hunits.  The constraints on the
lens models, and hence on $H_0$, can be improved by reducing
the standard errors in the lens galaxy position from $50$ mas
to $\sim10$ mas, reducing the uncertainties in the time
delays to $\sim0.5$ days, and constraining the lens mass
distribution using HST photometry and the fundamental plane.
In particular, the time delay ratio $\RABC\equiv\Delta\tau_{AC}/
\Delta\tau_{BA}$ may provide the best constraint on the mass
profile of the galaxy.
\end{abstract}


\bigskip

\centerline{\it To appear in The Astrophysical Journal}

\clearpage

\section{Introduction}

A general consensus is emerging from local studies that the Hubble
constant $H_0$ lies in the range $65$--$75$ \Hunits\ (e.g.\ Freedman,
Madore \& Kennicutt 1996), based primarily on Cepheid distances to
nearby galaxies combined with type Ia supernovae to reach regions of
pure Hubble flow (e.g.\ Riess, Press \& Kirshner 1996).  The resulting
age estimates are in weak conflict with the estimated ages of the oldest
stars for almost all cosmological models (e.g.\ Bolte \& Hogan 1995).
Measurements of time delays in gravitational lenses can directly
determine $H_0$ over cosmological distances (Refsdal 1964), thereby
avoiding the complicated calibration problems that plague local
distance estimates.  In a gravitational lens, the ray trajectories
of the multiple images have different geometric lengths and pass
through different parts of the gravitational potential, so the light
travel time is different for each image.  The light travel time is
inversely proportional to $H_0$, so combining observed time delays
between images with a model of the gravitational potential gives the
Hubble constant.  Unfortunately, application of this technique has
been slowed by the difficulties of determining the time delays and
of finding good models for the gravitational potential.

The first gravitational lens discovered, 0957$+$561 (Walsh et al.\
1979), was also the first lens for which a time delay was measured
(e.g.\ Vanderreist et al.\ 1992; Leh\'ar et al.\ 1992; Press, Rybicki
\& Hewitt 1992a,b; Schild \& Thomson 1995; Pelt et al.\ 1996; Kundi\'c
et al.\ 1996); the most recent measurement yielded the time delay
$\Delta\tau=417\pm3$ days at $95\%$ confidence (Kundi\'c et al.\ 1996).
The system has been modeled extensively (e.g.\ Young et al.\ 1980;
Borgeest \& Refsdal 1984; Narasimha, Subramanian \& Chitre 1984;
Greenfield, Roberts \& Burke 1985; Falco, Gorenstein \& Shapiro 1991;
Kochanek 1991b; Bernstein, Tyson \& Kochanek 1993; Grogin \& Narayan
1996).  Most recently, Grogin and Narayan (1996) considered a spherical
softened power-law model for the primary lens galaxy and an external
shear for the surrounding cluster.  The best-fit galaxy model had a
dark matter halo whose mass increases slightly faster than isothermal,
$M(r)\propto r^\al$ with $1.07<\al<1.18$ at $95\%$ confidence.  This
model gave $H_0 = \left(85_{-7}^{+6}\right) (1-\kappa)(\Delta\tau/1.1\
{\rm yr})^{-1}$ \Hunits, where $\kappa>0$ is an inherent degeneracy
due to the mean surface mass density of the cluster (Falco, Gorenstein
\& Shapiro 1985).  Kundi\'c et al.\ (1996) estimated $\kappa$ from the
observation of the cluster by Fischer et al.\ (1997) and used the model
of Grogin \& Narayan to infer $H_0=64\pm13$ \Hunits\ ($95\%$ confidence);
independently, Falco et al.\ (1997) used the velocity dispersion of the
galaxy to infer $H_0=62\pm7$ \Hunits.  We note that Grogin \& Narayan
(1996) briefly considered an approximate elliptical model of the galaxy,
but they fixed the ellipticity and position angle to match the observed
isophotes and did not explore the additional freedom in the models (and
hence in $H_0$) due to treating the galaxy and cluster as independent
sources of shear.

\begin{figure}[h]
	\plotone{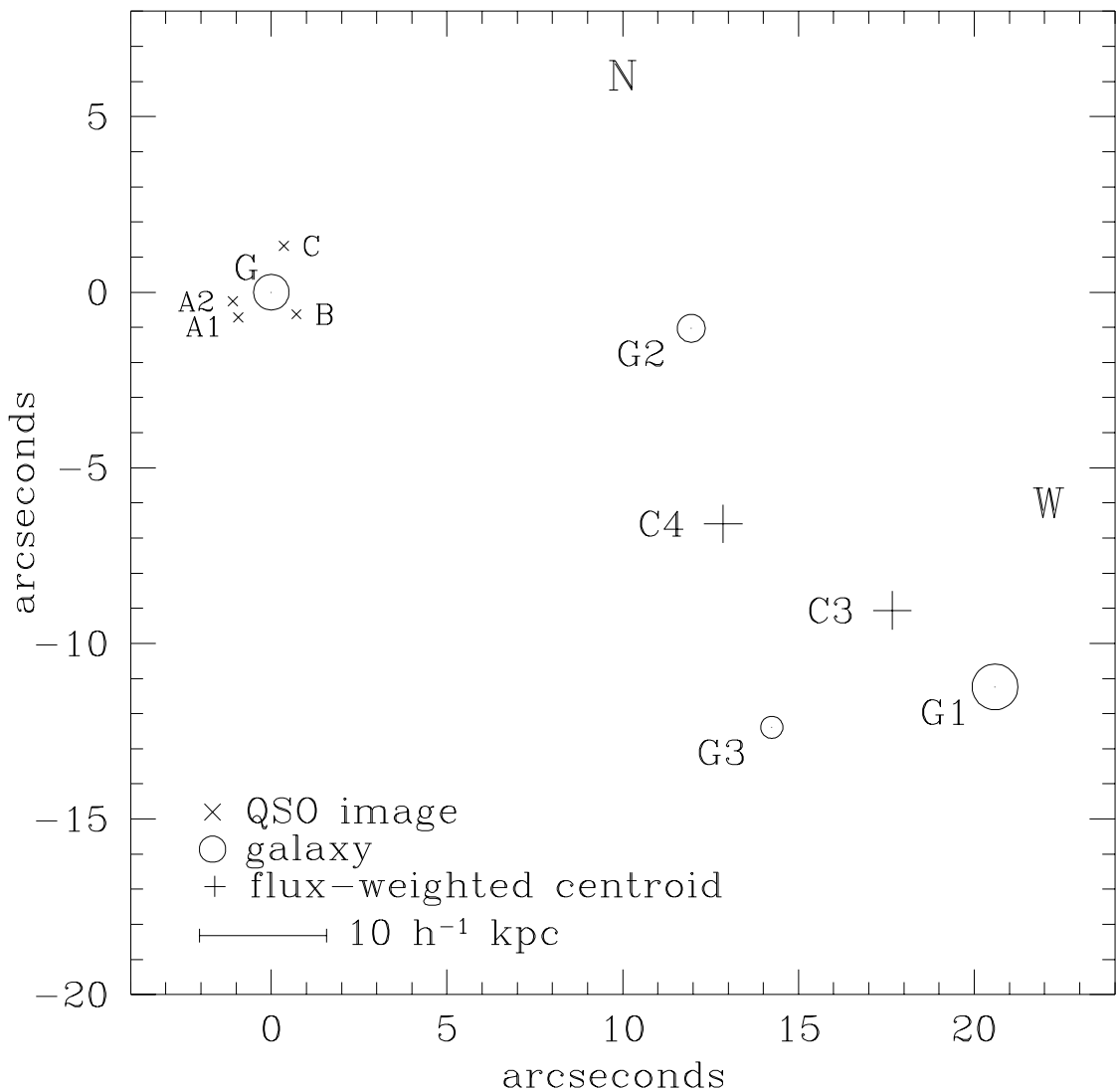}
	\caption{
Schematic diagram of the gravitational lens \1115\ and the nearby
group.  The crosses represent the four quasar images (K93).  The circles
represent the galaxies (Young et al.\ 1981), with the areas of the circles
indicating the relative fluxes.  The plusses represent (``C3'') the
flux-weighted centroid of the three group galaxies and (``C4'') the
flux-weighted centroid of all four galaxies.  The physical scale
indicated is for an $\Omega_0=1$ cosmology.
}
	\label{fig:schematic}
\end{figure}

Recently the four-image gravitational lens \1115\ (Weymann et al.\
1980; see \reffig{schematic}) became the second lens for which time
delays were measured.  Schechter et al.\ (1997) measured the time
delay between images $C$ and $B$ to be $\Delta\tau_{BC}=23.7\pm3.4$
days and the time delay between $C$ and mean of the close pair
$A=A_1+A_2$ to be $\Delta\tau_{AC}=9.4\pm3.4$ days, giving a time
delay ratio $\RABC\equiv\Delta\tau_{AC}/\Delta\tau_{BA}=0.7\pm0.3$.
Bar-Kana (1997), however, reanalyzed the data to show that including
the correlations in the photometric errors gives $\Delta\tau_{BC}=
25.0_{-1.7}^{+1.5}$ days and $\RABC=1.13_{-0.17}^{+0.18}$ (statistical),
with a $\sim0.2$ systematic uncertainty in $\RABC$ associated with
different assumptions about how to treat the photometric errors.

\1115\ is a promising candidate for combining time delays with lens
models to determine $H_0$, because four-image lens geometries can
constrain some aspects of lens models better than two-image lens
geometries (Kovner
1987; Kochanek 1991a).  \1115\ has been modeled extensively.  The
first models demonstrated that simple non-axisymmetric galaxy models
could qualitatively reproduce the image configuration (Young et al.\
1981; Narasimha, Subramanian \& Chitre 1982).  Further models showed
that an isothermal sphere or a point mass with a variety of quadrupole
structures could reproduce the image positions to within ground-based
observational errors, but that the observations were not
accurate enough to distinguish between the various monopole and
quadrupole forms (Kochanek 1991a).  More accurate observations with
the Hubble Space Telescope (Kristian et al.\ 1993, hereafter K93)
have provided better observational constraints.  Keeton, Kochanek \&
Seljak (1997a) showed that the quasar image positions and fluxes and
the galaxy position cannot be fit by models with a single shear axis
(such as an ellipsoidal galaxy or a circular galaxy with an external
shear), but can be fit by models allowing two independent shear axes
(such as an ellipsoidal galaxy with an external shear).  Schechter
et al.\ (1997) demonstrated that the nearby group of galaxies seen
by Young et al.\ (1981) can provide the required external shear.
In addition, Schechter et al.\ (1997) used their measurements of
the time delay with a simple lens model treating the galaxy and
group as singular isothermal spheres to infer $H_0=42$ \Hunits, with
a $14\%$ uncertainty from the time delay and with unknown uncertainties
from the lens model.  Their value for $H_0$ is in strong conflict with
local estimates of $H_0=65$--$75$ \Hunits, so it is crucial to
understand the systematic uncertainties in the lens models and their
effects on the inferred value of $H_0$.  In addition, if we hope to
use lensing to measure $H_0$ with precision comparable to the distance
ladder, then we must understand which future observations can reduce
the uncertainties in the lens models.

In this paper we examine the systematic uncertainties in the value of
$H_0$ inferred from \1115\ by exploring the uncertainties in the lens
models.  Since present observations of the galaxy and group do not
directly constrain their mass distributions and only weakly constrain
their positions, we postulate a wide range of plausible galaxy and group
properties.  Specifically, we model the galaxy as an ellipsoidal mass
distribution with a range of profiles including constant mass-to-light
ratio models, standard dark matter models, and more centrally-concentrated
models.  We model the group as a circular mass distribution with an
extended (isothermal) or concentrated (point mass) profile.  We then
study the ability of the data to constrain the models and show that
present constraints leave significant degeneracies in the models.  We
examine the implications of the models and degeneracies for $H_0$,
and discuss future observations that can break the degeneracies and
improve the constraints on $H_0$.  In \S2 we present the data and
methods used in the models.  In \S3 we consider a broad range of models
using only the primary lens galaxy and show that they cannot fit the
data.  In \S4 we study models adding the nearby group and show that
a variety of galaxy+group models give good fits.  We use a Bayesian
analysis with the most physically plausible models to quantify the
present systematic uncertainties in $H_0$.  In \S5 we illustrate
an independent way to break one degeneracy by using stellar dynamics
and the fundamental plane of elliptical galaxies to identify which
galaxy profiles are physically plausible.  In \S6 we summarize our
results for $H_0$ as well as the prospects for breaking the
degeneracies through better observations of the quasar fluxes, the
galaxy position and photometry, and the time delays.

\section{Data and Methods}

\1115\ consists of four point images of a $z_s=1.722$ quasar with
separations $\sim2\arcs$ surrounding a galaxy at redshift $0.295\pm
0.005$ (Weymann et al.\ 1980; K93; Angonin-Willaime et al.\ 1993).
There is a nearby group of galaxies (Young et al.\ 1981) at redshift
$0.304$ (Henry \& Heasley 1986).  A schematic diagram of the lens and
the galaxies is shown in \reffig{schematic}.  For reference, in an
$\Omega_0=1$ cosmology $1\arcs$ at the group redshift is $2.77 h^{-1}$
kpc.  Other cosmological scale factors used in the lensing analysis
are given in \reftab{cosmology} for several cosmologies.  Throughout
the text we use an $\Omega_0=1$ cosmology.

Computing $H_0$ from a gravitational lens requires time delays and a
lens model.  The time delays were taken from Bar-Kana's (1997) analysis
of the data of Schechter et al.\ (1997).  The $B-C$ time delay is the
longest and best-resolved time delay.  Bar-Kana (1997) found $\Delta
\tau_{BC}=25.0_{-1.7}^{+1.5}$ days, consistent with the value $23.7
\pm3.4$ days from Schechter et al.\ (1997), and this result depended
only weakly on assumptions about the photometric errors.  By contrast,
the other time delays and hence the time delay ratio $\RABC\equiv
\Delta\tau_{AC}/\Delta\tau_{BA}$ are less well determined.  Bar-Kana
(1997) found $\RABC=1.13_{-0.17}^{+0.18}$, in conflict with the value
$0.7\pm0.3$ from Schechter et al.\ (1997), and this result did depend
on assumptions about the photometric errors.  A range of assumptions
indicated that Bar-Kana's (1997) result for $\RABC$ was more robust
than the result of Schechter et al.\ (1997), and that systematic
uncertainties in $\RABC$ were at the level of $\sim0.2$ (Bar-Kana 1997).

To determine the lens model, we fitted lens mass distributions to
reproduce the quasar images positions and fluxes and the galaxy position
from the Hubble Space Telescope observations by K93 (see \reftab{data}).
The relative coordinates have an uncertainty of $5$ mas for the quasar
images and $50$ mas for the lens galaxy.  The quasar flux ratios are
less well determined because of the source variability, microlensing,
and extinction.  To account for this, we broadened the flux error bars
to $20\%$ to encompass the range of observed variability (see Keeton
\& Kochanek (1996) for a summary).  We also considered the effects on
our conclusions of making the flux error bars smaller (see \S6).  K93
estimated the upper limit on the flux of a faint central image to be
$1$--$2\%$ of the flux of the brightest image, and we included this
constraint by setting the limit on the flux of the central image to
be $0\pm2\%$ of the flux of the brightest image.

Time delays offer an independent constraint on lens models, but only
if more than one is known.  The first time delay is used to determine
$H_0$, and the rest are combined into $H_0$-independent ratios that
constrain the models.  However, because of the systematic uncertainties
in the time delays for \1115\ we used them only to determine $H_0$,
not to constrain the models.  For each model, we used the $B-C$ time
delay to compute $H_0$, which we express as $H_0=100\h\dtBC^{-1}$ \Hunits.  
We used the other known time delay in the ratio $\RABC$ only as a
qualitative check of the consistency of the models.  We did, though,
examine the effects of including the $\RABC$ constraint in the Bayesian
analysis of $H_0$ in \S4.2.

To determine whether a lens mass distribution is consistent with the
data, we studied the images it can produce.  The lensing properties
of a surface mass distribution $\Sigma$ are described by its lensing
potential $\psi$ determined from the two-dimensional Poisson equation
$\nabla^2\psi=2\Sigma/\Sigma_{cr}$ (e.g.\ Schneider, Ehlers \& Falco
1992), where the critical surface density for lensing in angular units
is
\begin{equation}
	\Sigma_{cr} = {c^2 \over 4\pi G}\, {D_l D_s \over D_{ls}}
	= 2.34h^{-1}\,\left[{D_l D_s \over 2r_H D_{ls}}\right] \times
		10^{11}\, M_\odot\ {\rm arcsec}^{-2}, \label{eq:sigcr}
\end{equation}
$r_H=c/H_0$ is the Hubble radius, and $D_l$, $D_s$, and $D_{ls}$ are
angular diameter distances to the lens, to the source, and between the
lens and the source, respectively.  Values of the distance ratio
are given in \reftab{cosmology}.  The potential $\psi$ describes the
mapping between the source and image planes through the lens equation
\begin{equation}
	\vec{u} = \vec{x} - \vec{\nabla}\psi(\vec{x}), \label{eq:lens}
\end{equation}
where $\vec{x}$ is an angular position in the image plane and
$\vec{u}$ is an angular position in the source plane.  A source
at $\vec{u}$ maps to images at $\vec{x}_i$ that are roots of the
lens equation.  The images are deformed by the magnification tensor
$M_{ij}$ where $ M_{ij}^{-1} = \delta_{ij} - {\partial^2 \psi /
\partial x_i \partial x_j} $.  Because surface brightness is
conserved, the total magnification of an image is the ratio of
the area of the magnified image to the area of the source; so the
magnification factor is $M=\det(M_{ij})$.  The ray trajectories of
the images have different geometric lengths and pass through different
parts of the gravitational potential, so the light travel time is
different for each image; the time at image position $\vec{x}$ is
\begin{equation}
	\tau(\vec{x}) = {1+z_l\over c}\,{D_l D_s\over D_{ls}}\,\left[
		{1\over2}\biggl(\vec{x}-\vec{u}\biggr)^2 - \psi(\vec{x})\right],
		\label{eq:tau}
\end{equation}
and the time delay between an image at $\vec{x}_i$ and an image at
$\vec{x}_j$ is $\Delta\tau_{ij}=\tau(\vec{x}_i)-\tau(\vec{x}_j)$.  
Note that $\tau$ factors into a piece that depends on the lens model
times a piece that depends on cosmology and scales as $H_0^{-1}$ (see
\reftab{cosmology} for values in different cosmologies).  If we let
$\tilde D=D/r_H$ be an angular diameter distance scaled by $r_H$ (and
hence independent of $H_0$), then we can use \refeq{tau} to find $H_0$
from a lens model and time delay,
\begin{equation}
	H_0 = {1+z_l\over\Delta\tau_{ij}}\,{\tilde D_l \tilde D_s \over \tilde D_{ls}}
		\left[{1\over2}\biggl(|\vec{x}_i-\vec{u}|^2-|\vec{x}_j-\vec{u}|^2\biggr)
			-\biggl(\psi(\vec{x}_i)-\psi(\vec{x}_j)\biggr)\right].
		\label{eq:H0}
\end{equation}
From \reftab{cosmology}, changing the cosmology can change the inferred
value of $H_0$ by up to $7\%$.  

We modeled the primary lens galaxy with ellipsoidal surface mass
densities of the form
\begin{equation}
	\Sigma = \Sigma(m^2) \qquad {\rm where}\qquad
	m^2 = r^2 \biggl(1+\ep\cos2(\th-\te)\biggr),
\end{equation}
$\ep$ is a natural ellipticity parameter such that the axis ratio is
$(1-\ep)^{1/2}/(1+\ep)^{1/2}$, and $\te$ is the orientation angle of the
major axis.  We will quote $\te$ as a standard position angle measured
North through East.  We considered two classes of models.  First, we used
the de Vaucouleurs (1948) model as the prototypical constant mass-to-light
ratio ($M/L$) model for early-type galaxies, with surface mass density
\begin{equation}
	2{\Sigma \over \Sigma_{cr}} = {b\over r_e}\,
		{\exp\left\{-k\,(m^2/r_e^2)^{1/8}\right\} \over
		\int_0^\infty dv\, v\, \exp\left\{-k\,v^{1/4}\right\} },
\end{equation}
where $k=7.67$, $r_e$ is the effective (or half-light) radius, and
$b$ is the deflection scale such that the total mass is
\begin{equation}
	M_{tot} = {b r_e \over \sqrt{1-\ep^2}}\, {c^2 \over 4G}\,
		{D_l D_s \over D_{ls}}
	= 7.36 h^{-1} {b r_e \over \sqrt{1-\ep^2}}\,
		\left[ {D_l D_s \over 2r_H D_{ls}} \right] \times 10^{11}\, M_\odot
\end{equation}
for $b$ and $r_e$ in arcseconds.  Second, we used softened power-law
models with surface mass density
\begin{equation}
	2{\Sigma \over \Sigma_{cr}} = { b^{2-\al} \over
		\left(s^2 + m^2\right)^{1-\al/2} },
\end{equation}
where $\al$ is the power law exponent such that $M(r)\propto r^\al$
asymptotically, $s$ is a core radius, and $b$ is the deflection scale.
For $\al=1$ it is a softened isothermal model, whose lensing properties
have been studied by Kassiola and Kovner (1993) and Kormann, Schneider
\& Bartelmann (1994).  For $\al=0$ it is a modified Hubble model (Binney
\& Tremaine 1987).  We also studied the more centrally-concentrated
$\al=-1$ model.  For $\alpha=1,0,-1$ we found analytic expressions for
the deflection and the magnification, and for other values of $\alpha$
we used numerical integrals.

We modeled the group as a single halo in which the observed galaxies
are embedded.  For simplicity we used only circular mass distributions.
We considered a standard dark matter model treating the group as a
singular isothermal sphere (SIS), and we studied the effects of making
it more centrally concentrated by considering the limit of a point
mass.  The lensing potential for a point mass at the origin is
$\psi(r) = b^2\ln r$.

For a given lens mass distribution, we solved the lens equation
(\ref{eq:lens}) to map a source at $\vec{u}$ to its images at $\vec{x}_i$.
We varied the position and flux of the source and the parameters of the
mass distribution using the ``amoeba'' downhill simplex method (Press
et al.\ 1992) to minimize the residuals in the image plane.  We included
in the $\chi^2$ the constraints from the quasar positions and fluxes and
the galaxy position.  The galaxy position and the four quasar positions
and fluxes provide $14$ constraints.  A galaxy model with $r_e$ or
$(s,\al)$ fixed has five free parameters ($\vec{x}_{gal}$, $\ep$, $\te$,
and $b$).  A group model has one parameter ($\bgrp$) if the position is
fixed and three parameters if the position is variable.  The source has
three parameters ($\vec{u}$ and the flux).  Thus a lens model using an
isolated lens galaxy with fixed $r_e$ or $(s,\al)$ has $N_{dof}=6$ and
a lens model using the galaxy plus a movable group has $N_{dof}=3$.

\section{Results: Isolated Galaxy}

We first considered simple lens models using only the primary lens
galaxy.  Previous studies showed that the K93 data cannot be fitted
by an isolated lens galaxy if the galaxy is represented by a point
mass or a singular isothermal mass distribution (Keeton et al.\ 1997;
Schechter et al.\ 1997).  To see whether some other isolated galaxy
could fit the data, we considered a range of softened power-law
models, as well as de Vaucouleurs constant $M/L$ models.  For all
models we used an ellipsoidal mass distribution as the source of
shear.

\begin{figure}[h]
	\plotone{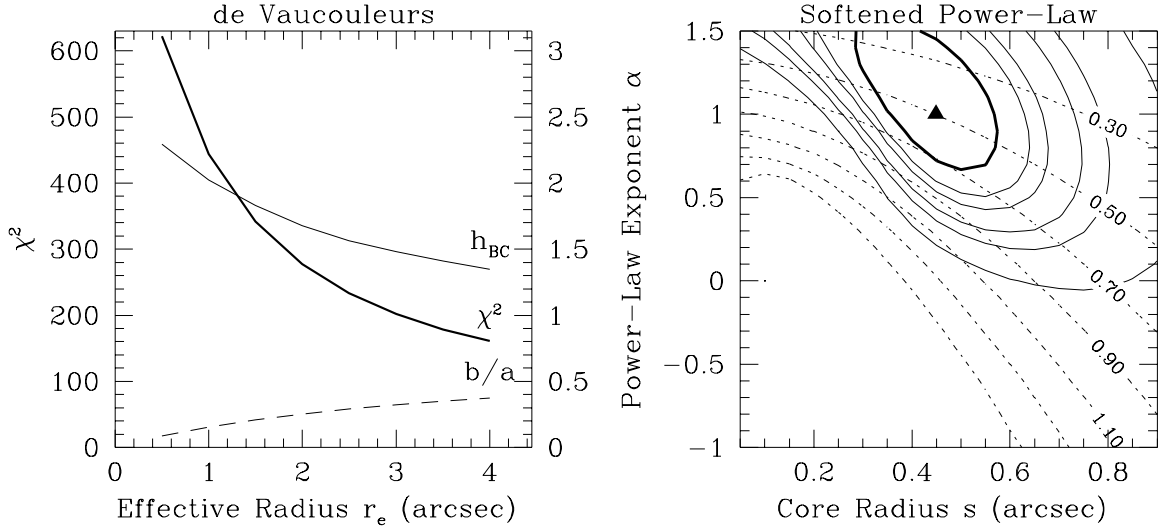}
	\caption{
Results for models using only the primary lens galaxy.  The galaxy is
represented by an ellipsoidal mass distribution with the specified
density profile.  Each model with fixed $r_e$ or $(s,\al)$ has $N_{dof}=6$.
{\it Left\/}:  de Vaucouleurs profile.  The heavy solid line is the
$\chi^2$ using the axis scale on the left.  The light solid line is
the Hubble constant expressed as $H_0=100\h\dtBC^{-1}$ \Hunits, and
the dashed line is the axis ratio of the galaxy; both use the axis
scale on the right.
{\it Right\/}:  Softened power-law profile.  The solid lines are contours
of $\chi^2$ drawn at $\Delta\chi^2=2.30$, $4.61$, $6.17$, $9.21$, $11.8$,
and $18.4$, the $1\sigma$, $90\%$, $2\sigma$, $99\%$, $3\sigma$, and
$99.99\%$ confidence levels for two parameters.  The dotted lines are
contours of $\h$.  The best-fit model (marked with a triangle) is at
$s=0\parcs449$, $\al=1.002$ and has $\chi^2=46.9$ and $\h=0.50$.
}
	\label{fig:nogrp}
\end{figure}

For the de Vaucouleurs models, there are no observational estimates
of the effective radius $r_e$, so we tabulated results as a function
of $r_e$ (see \reffig{nogrp}).  To estimate the plausible range of
$r_e$, we note that an $L_*$ galaxy has an effective radius of
$(4\pm1) h^{-1}$ kpc (Kormendy \& Djorgovski 1989; Rix 1991), which
corresponds to $1\parcs4$ at the redshift of \1115.  In addition, in
\S5 we use stellar dynamical arguments to estimate that $r_e$ lies in
the range $1\parcs0 \lesssim r_e \lesssim 4\parcs0$.  With this range
of effective radii, no de Vaucouleurs model gives an acceptable fit.
The $\chiN$ is $68$ at $r_e=1\parcs5$ ($4.2h^{-1}$ kpc) and is even
larger at smaller effective radii.  Although $\chiN$ decreases for
larger $r_e$, it is still $18$ at $r_e=10\arcs$ ($28h^{-1}$ kpc).
All physical parameters inferred from the models are unreasonable:
for $r_e=0\parcs5$--$4\parcs0$, the models require a very flattened
galaxy ($b/a\sim0.1$--$0.4$) and imply a huge Hubble constant
($\h\sim2.3$--$1.3$).  Moreover, the ratio of the model $A_1-C$
and $B-A_1$ time delays is $\RABC\sim5.4$--$2.0$, compared with
$0.7\pm0.3$ for the time delays of Schechter et al.\ (1997) or
$1.13\pm0.2$ for the time delays of Bar-Kana (1997).  A simple
constant $M/L$ galaxy cannot describe \1115.

For the softened power-law models, we tabulated the results in the
plane of the core radius $s$ and the power-law exponent $\al$ (see
\reffig{nogrp}).  Generally the softened power-law profiles fit much
better than the de Vaucouleurs profile but still do not give a good
fit.  The best-fit model has $\chiN=11.7$, is nearly isothermal
($\al=1.002$), and has a large core radius ($s=0\parcs45=1.2h^{-1}$
kpc).  This model implies an axis ratio $b/a=0.75$, a Hubble constant
$\h=0.50$, and a time delay ratio $\RABC=1.44$ that is marginally
consistent with Bar-Kana (1997) given his systematic uncertainties.
As $s$ or $\al$ decreases, the models become more centrally-concentrated
and imply larger values for $H_0$; this is because $H_0$ scales as
$1-\kappa_0$ where $\kappa_0$ is the convergence at the critical radius
(Falco et al.\ 1985), and the models with steeper profiles have a smaller
$\kappa_0$.  The models are not tightly constrained, with the $1\sigma$
region allowing core radii $0\parcs28\,(0.8h^{-1}\,\kpc)\lesssim s
\lesssim 0\parcs55\,(1.5h^{-1}\,\kpc)$, and power-law exponents
$0.7\lesssim \al \lesssim1.6$.  Profiles steeper than the modified
Hubble profile ($\al=0$) are ruled out at better than $99.9\%$
confidence.  Over the $1\sigma$ region, $\RABC$ is well constrained
and varies by only $\sim0.1$, but $\h$ is poorly constrained and
varies from $0.2$ to $0.7$.  For reference, the singular isothermal
ellipsoid model has $\chiN=25$, $b/a=0.50$, $\h=0.94$, and $\RABC=1.85$.

It is surprising that the models require such a large core radius
even for the isothermal profile.  We expect a small core radius for
two reasons.  First, observations of galactic cores show that the
luminosity densities of elliptical galaxies do not have flat cores but
instead have central cusps (Gebhardt et al.\ 1996).  Second, it is
generally argued that small core radii are needed to fit lens data,
both because almost all known lenses lack a central image, and because
stellar dynamics, lens statistics, and other lens models are consistent
with isothermal mass distributions with small core radii (Narasimha,
Subramanian \& Chitre 1986; Kassiola \& Kovner 1993; Wallington \&
Narayan 1993; Kochanek 1993, 1995, 1996; Grogin \& Narayan 1996).
Contrary to these expectations, \1115\ cannot be fit by an isolated
lens galaxy with a small core radius.  It is difficult, though, to draw
strong conclusions from this result because the ``best-fit'' model is
still not a good fit ($\chiN=11.7$), while once we include the
perturbation from the group we will find good fits ($\chiN\lesssim1$)
with galaxies that do have a small core radius.

We conclude that \1115\ cannot be fit by an isolated lens galaxy, and
thus it is not reasonable to use such a model to estimate $H_0$.

\section{Results: Galaxy + Group}

Keeton et al.\ (1997a) found that while \1115\ could not be fit by
an isolated isothermal galaxy, it could be well fit by an ellipsoidal
galaxy with an independent external shear.  Schechter et al.\ (1997)
pointed out that the group seen by Young et al.\ (1981) was correctly
located to be the source of the shear, and by modeling the galaxy and
the group as two singular isothermal spheres they inferred $H_0=42$
\Hunits.  This value of $H_0$ is unexpectedly low, but it is difficult
to judge its significance because Schechter et al.\ (1997) did not
explore the systematic uncertainties in the lens models and their
effects on the inferred $H_0$.  In order to understand the systematic
effects, we now examine the galaxy+group models in detail.  We explore
the freedom in the models (and hence in $H_0$) and discuss the prospects
for better constraining the models with future observations.  First we
consider models with the group position fixed to examine the effects
of the profiles of the galaxy and the group, and then we consider models
in which the group position is allowed to vary.

\subsection{Fixed group}

The group mass distribution is described by a position and any
parameters associated with the radial profile.  Let $\dgrp$ be the
distance from the primary lens galaxy to the group, and $\bgrp$ be
the critical radius of the group.  To lowest order the group is a
perturbation that can be characterized by its convergence $\kgrp$
and shear $\ggrp$ together with weaker nonlinear terms (Falco et
al.\ 1985).  For a singular isothermal sphere (SIS) group $\ggrp=
\kgrp=\bgrp/2\dgrp$, while for a point mass group $\ggrp=\bgrp^2/
\dgrp^2$ and $\kgrp=0$.  Since $\ggrp$ scales with $\bgrp/\dgrp$,
we expect that requiring an external shear $\ggrp\sim10\%$ (Kochanek
1991a; Keeton et al.\ 1997a; Schechter et al.\ 1997) will produce
(to lowest order) a degeneracy between the group's position and its
mass.  The non-linear terms may break the degeneracy (see \S4.2),
but for simplicity we first consider models with the group position
fixed.  In this way we can focus on the effects of different profiles
for the galaxy and the group.

The three group galaxies and the lens galaxy are at essentially the
same redshift and are probably physically related, so we placed the
group at the flux-weighted centroid of all four galaxies (C4 in
\reffig{schematic}).  The Gunn $r$ magnitudes of the three group
galaxies were taken from Young et al.\ (1981); the $r$ magnitude
of the primary lens galaxy was estimated from the F785LP magnitude
(K93) by using the galaxy evolution models of Bruzual \& Charlot
(1993) to compute the color of an E/S0 galaxy at a redshift of
$0.3$ (see Keeton, Kochanek \& Falco 1997b for details).  We found
the flux centroid to be $\dgrp=14\parcs5$ from the primary lens
galaxy at a position angle $\tgrp=-117\pdeg1$ (North through East).

\begin{figure}[h]
	\plotone{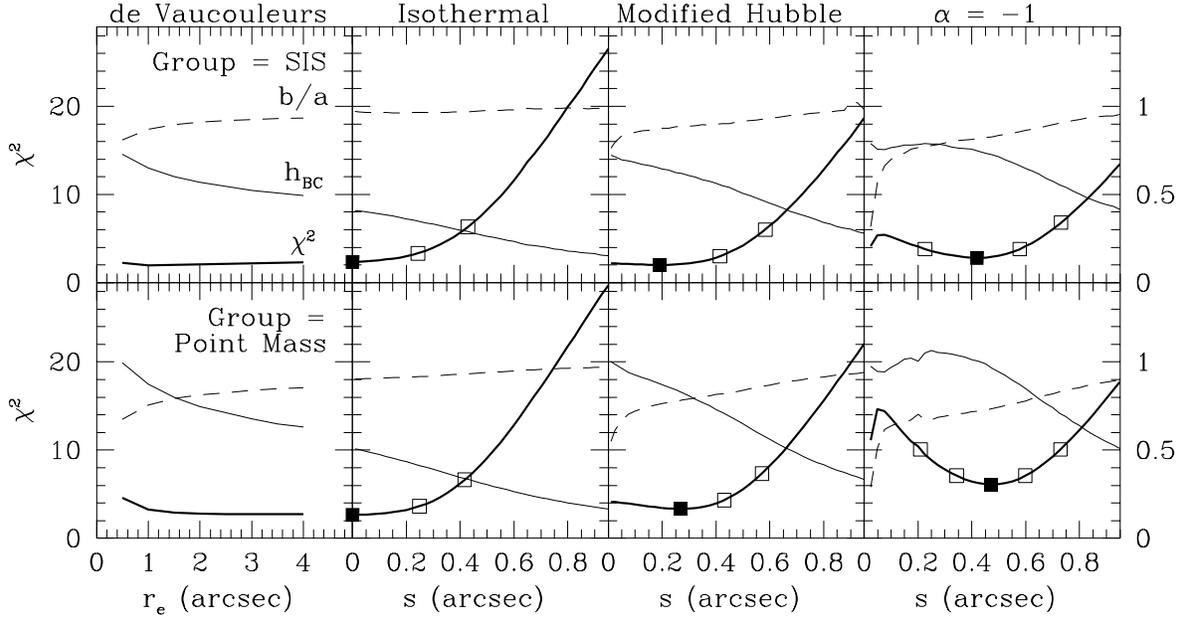}
	\caption{
Results for models using the primary lens galaxy plus the nearby group.
The group position is fixed at the flux-weighted average of the four
galaxies (C4 in Figure 1).  The primary lens galaxy is represented by
an ellipsoidal mass distribution and the group is a represented by a
circular mass distribution.  {\it Top\/}:  the group is represented by
an SIS.  {\it Bottom\/}:  the group is represented by a point mass.
The heavy solid line is the $\chi^2$ using the axis scale on the left;
each model with fixed $r_e$ or $s$ has $N_{dof}=5$.  The filled and open
points indicate the best fit and the $\Delta\chi^2=1$ and $\Delta\chi^2=4$
limits.  The light solid line is $\h$ and the dashed line is the axis
ratio of the galaxy, both using the axis scale on the right.
}
	\label{fig:fixgrp}
\end{figure}

\reffig{fixgrp} shows the results for de Vaucouleurs, softened
isothermal ($\al=1$), modified Hubble ($\al=0$), and $\al=-1$ models
of the primary lens galaxy with either an SIS or a point mass group,
and \reftab{results} summarizes the $\chi^2$ and the physical
parameters for the best fits.  All eight classes of models give good
fits, with $\chiN<1$ for all but the model with an $\al=-1$ galaxy
and a point mass group.  Since no model with an isolated galaxy could
do better than $\chiN \sim10$, while numerous models with the galaxy
supplemented by the group produce $\chiN\lesssim1$, we confirm the
results of Keeton et al.\ (1997a) and Schechter et al.\ (1997) that
including the group is crucial to obtaining a good fit.  We note that
the best model of Schechter et al.\ (1997), which treated the galaxy
and group and singular isothermal spheres, gave $\chiN=5$.  Our models
give $\chiN<1$ despite the addition of the flux constraints because we
allowed the galaxy position to vary, and because we included a second
shear by allowing the galaxy to have an ellipticity.  As we noted
in Keeton et al.\ (1997a), the key to fitting many of the quadruple
lenses is having two independent shear axes.

The values for $\h$ implied by these models vary significantly with
the galaxy and group profiles (see \reftab{results}).  Qualitatively,
the variation makes sense:  $H_0$ scales with $1-\kappa_0$ (Falco et
al.\ 1985), and the convergence $\kappa_0$ at the critical radius
decreases as the galaxy or group becomes centrally concentrated.
Unfortunately, the data cannot constrain the profiles for two reasons.
First, although the four-image configuration constrains the total mass
within the critical radius, it does not constrain the distribution of
the mass (Kochanek 1991a; Wambsganss \& Paczy\'nski 1994).  Second,
although the quasar images constrain the shear from the group, they
do not constrain its mass or convergence.  As a result, we can have
a wide range of galaxy and group profiles that are all consistent at
the $1\sigma$ level, and by using different profiles we can produce
$\h$ anywhere from $0.4$ to more than $0.8$ without significantly
changing the goodness of fit.  As we show in \S5, though, not all of
the profiles corresponds to galaxies that are physically plausible.
The centrally concentrated models that allow the higher values for
$H_0$ correspond to galaxies that do not lie on the fundamental
plane of elliptical galaxies.  Restricting attention to plausible
dark matter and constant $M/L$ models restricts the range of $\h$
to $0.4 \lesssim \h \lesssim 0.7$.

\subsection{Movable group}

\begin{figure}[h]
	\plotone{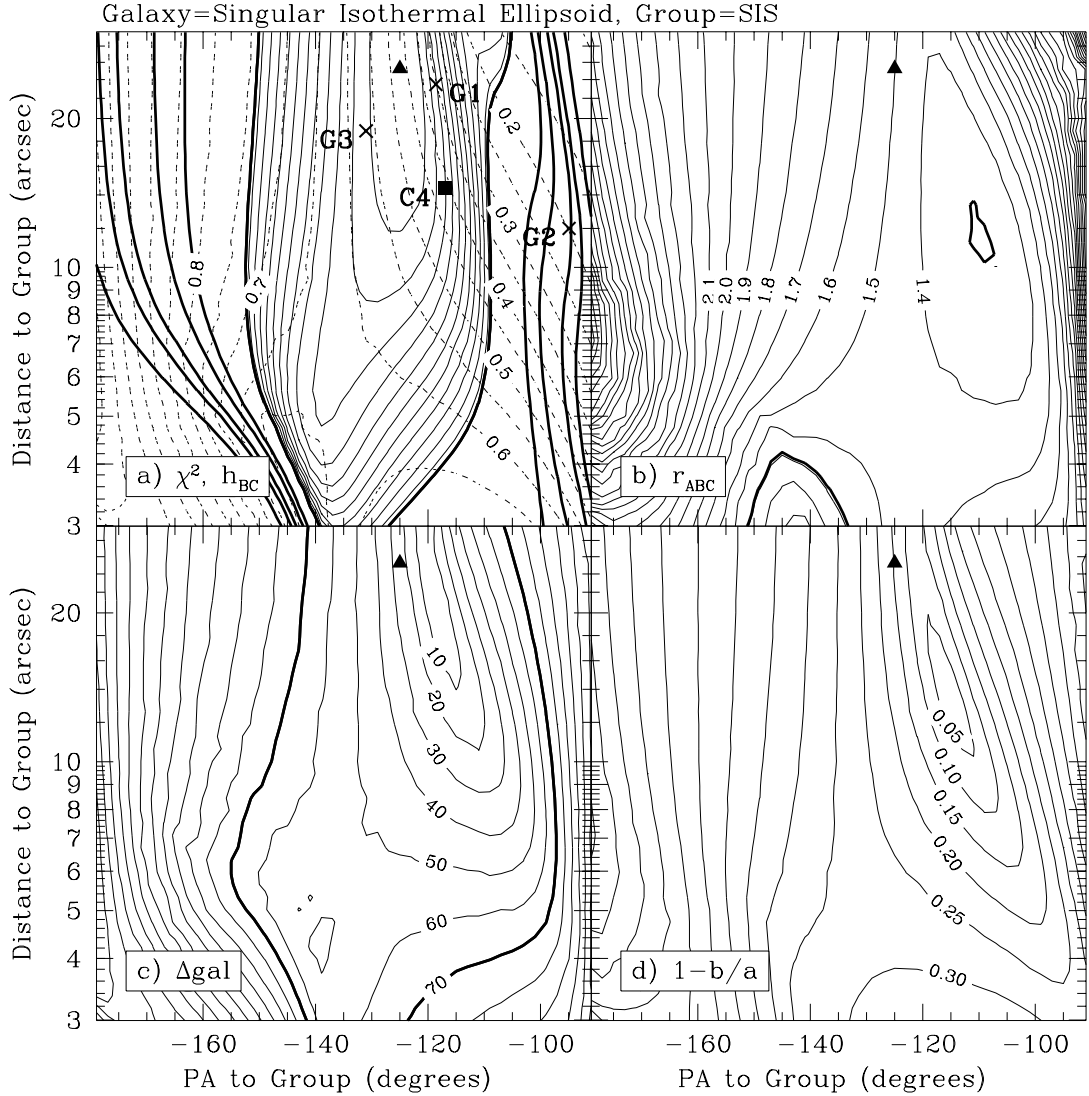}
	\caption{
Results for models representing the primary lens galaxy as a singular
isothermal ellipsoid and the group as an SIS.  With the group position 
fixed a model has $N_{dof}=5$.
(a) The solid lines are contours of $\chi^2$ drawn at $\Delta\chi^2=0.2$,
$0.4$, $\ldots$ (light) and at $\Delta\chi^2=2.30$, $4.61$, $6.17$, $9.21$,
$11.8$, and $18.4$ (heavy), the $1\sigma$, $90\%$, $2\sigma$, $99\%$,
$3\sigma$, and $99.99\%$ confidence levels for two parameters.  The
dashed lines are contours of $\h$.  The positions of the group galaxies
and of the flux-weighted centroid (C4 in Figure 1) are indicated.
(b) Contours of the time delay ratio $\RABC$.  The heavy contours indicate
the $1\sigma$ range $1.13_{-0.17}^{+0.18}$ from Bar-Kana (1997).
(c) Contours of the distance $\Delta_{gal}=|\vec{x}_{mod}-\vec{x}_{obs}|$
(in mas).  The heavy contour indicates the range allowed by the Kristian
et al.\ (1993) error bars, $\Delta_{gal}\le\radical"270370{2}\times50$ mas.
(d) Contours of the galaxy ellipticity $1-b/a$.  In the upper right corner
the galaxy is oriented approximately North--South, and everywhere else it
is approximately East--West.
The best-fit model (marked with a triangle) is at $\dgrp=25\parcs2$ and
$\tgrp=-125^\circ$ and has $\chi^2=1.77$, $\h=0.47$, $\RABC=1.50$,
$\Delta_{gal}=25$ mas, and $b/a=0.85$.
}
	\label{fig:varygrp1}
\end{figure}

\begin{figure}[h]
	\plotone{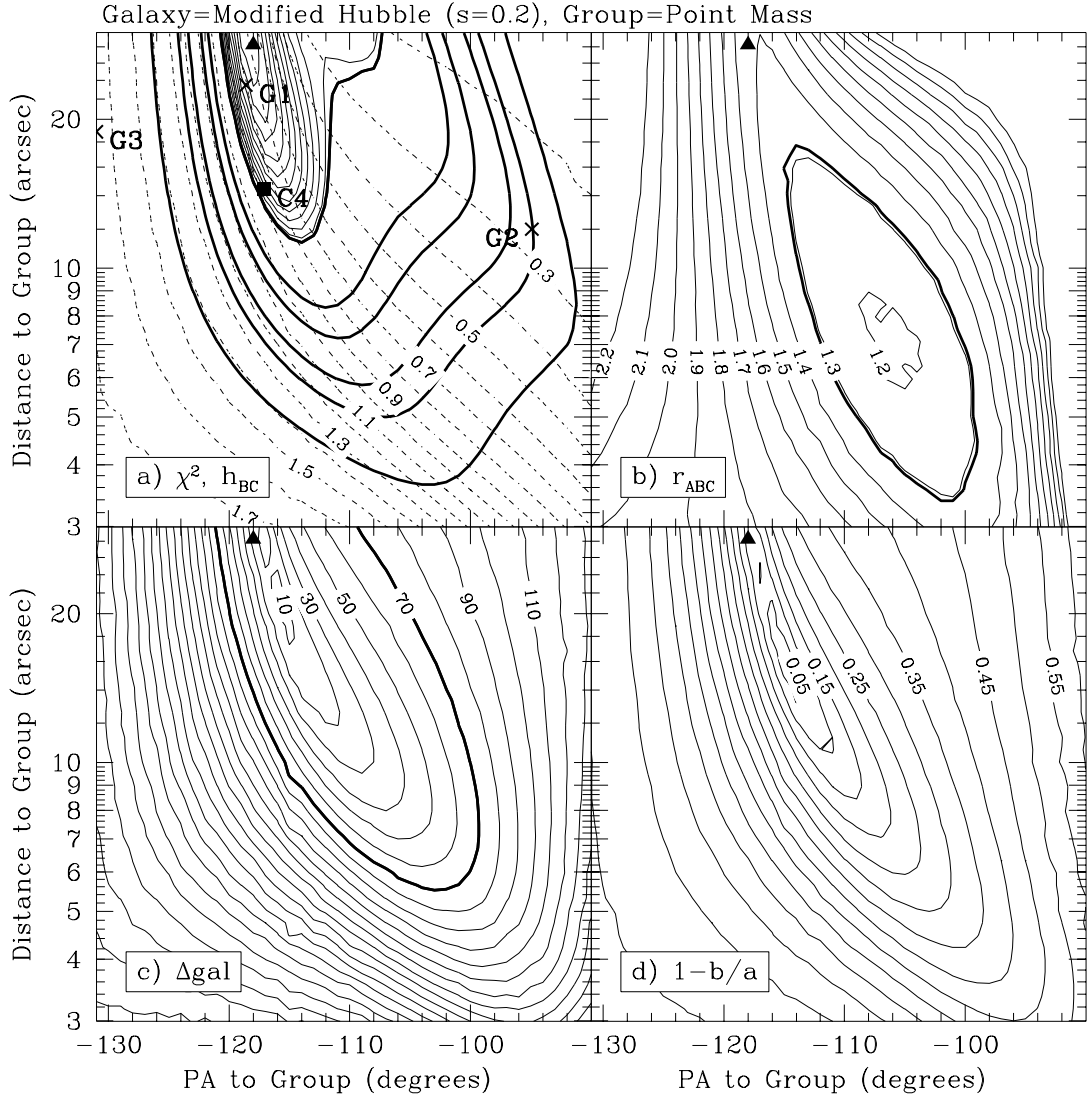}
	\caption{
Results for models representing the primary lens galaxy as a modified
Hubble model with core radius $s=0\parcs2$ ($0.55h^{-1}$ kpc) and the
group as a point mass.  The panels are the same as in Figure 4, except
that the range of position angles is smaller.  Again in (d) the galaxy
is oriented approximately East--West in the lower left and North--South
in the upper right.
The best-fit model (marked with a triangle) is at $\dgrp=28\parcs3$ and
$\tgrp=-118^\circ$ and has $\chi^2=1.70$, $\h=0.67$, $\RABC=1.43$,
$\Delta_{gal}=16$ mas, and $b/a=0.92$.
}
	\label{fig:varygrp2}
\end{figure}

While fixing the group at the flux centroid C4 allowed us to isolate
the effects of the profiles of the galaxy and group, there is no
strong observational evidence requiring it.  Hence the group position
is an additional systematic uncertainty whose effects on $H_0$ must be
examined.  We studied the effects of the group position for two galaxy
models:  a dark matter (isothermal) model and a constant $M/L$ model.
For the constant $M/L$ model we used a modified Hubble profile because,
although it is not as good a representation of galaxy luminosity profiles
as the de Vaucouleurs profile, it does have an analytic deflection
formula.  We neglected the more centrally concentrated galaxy models
because, as we show in \S5, they require galaxies that are unphysical.
For the group we again used the SIS and point mass models.  Physically
we expect the group to be described by an isothermal dark matter halo,
but by including the point mass group we can examine the effects of
making the group more concentrated.  Since these two galaxy profiles
and two group profiles span the expected range of mass profiles, they
should span the expected range of $H_0$.  We examined all four
combinations of profiles; representative results are shown in
\reffigs{varygrp1}{varygrp2}, where \reffig{varygrp1} shows the extended
profiles (isothermal galaxy and SIS group), and \reffig{varygrp2} shows
the concentrated profiles (modified Hubble galaxy and point mass group).

The singular isothermal ellipsoid (dark matter) galaxy permits good fits
for a wide range of group positions (see \reffig{varygrp1}).  With an SIS
group, the best-fit model is at $\dgrp=25\parcs2$ and $\tgrp=-125\deg$,
has $\chi^2=1.77$ for $N_{dof}=3$, and implies $\h=0.47$.  But at the
$1\sigma$ level, $\tgrp$ can range from $-150\deg$ to $-110\deg$ and
$\dgrp$ can be as small as a few arcseconds.  The Hubble constant
varies considerably over the allowed region, from $\h=0.2$ for distant
groups with $\tgrp\simeq -113\deg$, to $\h=0.7$ for nearby groups or
for distant groups with $\tgrp\simeq -145\deg$.

Note that there is no formal upper limit on $\dgrp$ from the $\chi^2$,
because for a distant group the non-linear terms are weak and the
perturbation is equivalent to a simple external shear, which we know
fits well (Keeton et al.\ 1997a).  However, the mass of the group
increases with its distance ($M_{grp}\propto\bgrp\dgrp\sim\dgrp^2$),
so requiring that the group have a reasonable mass can constrain its
position.  It is convenient to study the mass of the group in terms of
its mass-to-light ratio ($M/L$), and it is convenient to express the
$M/L$ of the group in units of the $M/L$ of the galaxy because this
ratio is independent of assumptions about the cosmology and about the
$K$ and evolutionary corrections to the luminosities.  The models with
the group in a band from the centroid C4 to the best-fit group position
have $5 \lesssim (M/L)_{grp}/(M/L)_{gal} \lesssim 25$.  The galaxy
mass within the critical radius (radius $r_0=1\parcs14=3.16h^{-1}$
kpc) is $1.22\times10^{11} h^{-1}\ M_\odot$, with a $5\%$ uncertainty
due to the position of the group, a $10\%$ uncertainty (upwards) due
to making the group more centrally concentrated, and a $7\%$
uncertainty (upwards) due to the cosmological model.  Using the K93
F785LP central aperture magnitude together with color, $K$, and
evolutionary corrections from the galaxy evolution models of Bruzual
\& Charlot (1993), we estimated the present $B$ magnitude of the galaxy
to be $-19.2$ mag for an $\Omega_0=1$ cosmology (see Keeton et al.\
1997b for details), although the effective aperture of the K93
magnitude is unclear.  This gives $(M/L)_{gal}\simeq18$, which is
consistent with other lens models (Keeton et al.\ 1997b) and lens
statistics (Kochanek 1993, 1996; Maoz \& Rix 1993).  This in turn
implies $90 \lesssim (M/L)_{grp} \lesssim 450$, which is consistent
with observed group mass-to-light ratios, e.g.\ $150 \lesssim M/L
\lesssim 350$ (Ramella, Pisani \& Geller 1997).  It is difficult at
present to apply strong constraints from this reasoning, but it is
clear that the group positions that give good fits also give reasonable
group masses and that the group should not be much closer or much
farther away.

\reffig{varygrp1} shows that at present the constraints on $H_0$ are
relatively weak.  Nevertheless, the models imply several physical
properties that may be better constrained by future observations.
First, the galaxy positions differ at the level of tens of mas.  While
the K93 position is well within the range of good fits, the $50$ mas
error bars are too broad to rule out many models.  Reducing
the error bars on the galaxy position to $10$ mas or better would
greatly improve the constraints on the models.  Second, the galaxy
axis ratios vary.  The galaxy is strictly circular if the group is
at $\dgrp=13\parcs5$, $\tgrp=-115\deg$ (which reproduces the double
SIS model of Schechter et al.\ 1997).  For other group positions
the galaxy's ellipticity and orientation adjust so the combined shear
from the galaxy and group gives a good fit.  The ellipticity needs
to vary by only a few times $0.05$ in order to produce good fits
for a wide variety of group positions.  Finally, the time delays
vary.  Although it was not included as a constraint, the ratio
$\RABC$ is marginally consistent with the value $1.13\pm0.2$ of
Bar-Kana (1997) given his systematic uncertainties of $\sim0.2$.
Note that no models produce a time delay ratio consistent with the
value $\RABC=0.7\pm0.3$ of Schechter et al.\ (1997).  The time delay
ratio could provide a strong constraint on the models if the systematic
effects in the observed ratio were understood and if the uncertainties
were reduced to $\sim0.05$.  This would require roughly a factor of
four reduction in the uncertainties in the time delays, or uncertainties
of $\lesssim0.5$ days.  We note also that the variation of the $A_2-A_1$
time delay with group position differs from that of the $A-C$ and $B-A$
time delays, so this time delay could provide an independent constraint.
However, $\Delta\tau_{A_2 A_1}=5\pm1$ hours over the $1\sigma$ region
so this time delay would need to be known at the level of $\sim0.5$ hour
in order to constrain the models.

The results are similar for the modified Hubble model (constant $M/L$)
galaxy, except that good fits are limited to a narrower range of group
positions (see \reffig{varygrp2}).  With a fixed core radius $s=0\parcs2$
($0.55h^{-1}$ kpc) and a point mass group, the best-fit model is at
$\dgrp=28\parcs3$ and $\tgrp=-118\deg$, has $\chi^2=1.70$ for $N_{dof}=3$,
and implies $\h=0.67$.  At the $1\sigma$ level,
$\tgrp$ is restricted to $-120\deg \lesssim \tgrp \lesssim -112\deg$
and the group can be no closer than $\dgrp=11\arcs$.  As with the
isothermal galaxy, the Hubble constant varies considerably over the
$1\sigma$ region, ranging from $0.3$ to $0.9$.  However, again physical
parameters including the galaxy position, the galaxy axis ratio, and
the time delays may be constrained by further observations to improve
the constraints on $H_0$.  In particular, note that the modified Hubble
galaxy predicts a time delay ratio that is closer to the observed value
of Bar-Kana (1997) than the ratio predicted by the isothermal galaxy.
Although it is not shown, the predicted time delay ratio is essentially
independent of the profile of the group.  Thus if Bar-Kana's (1997) value
remains valid as its uncertainties are reduced, then it may provide
the strongest probe of the galaxy profile.  We will return to this point
below.

\begin{figure}[h]
	\plotone{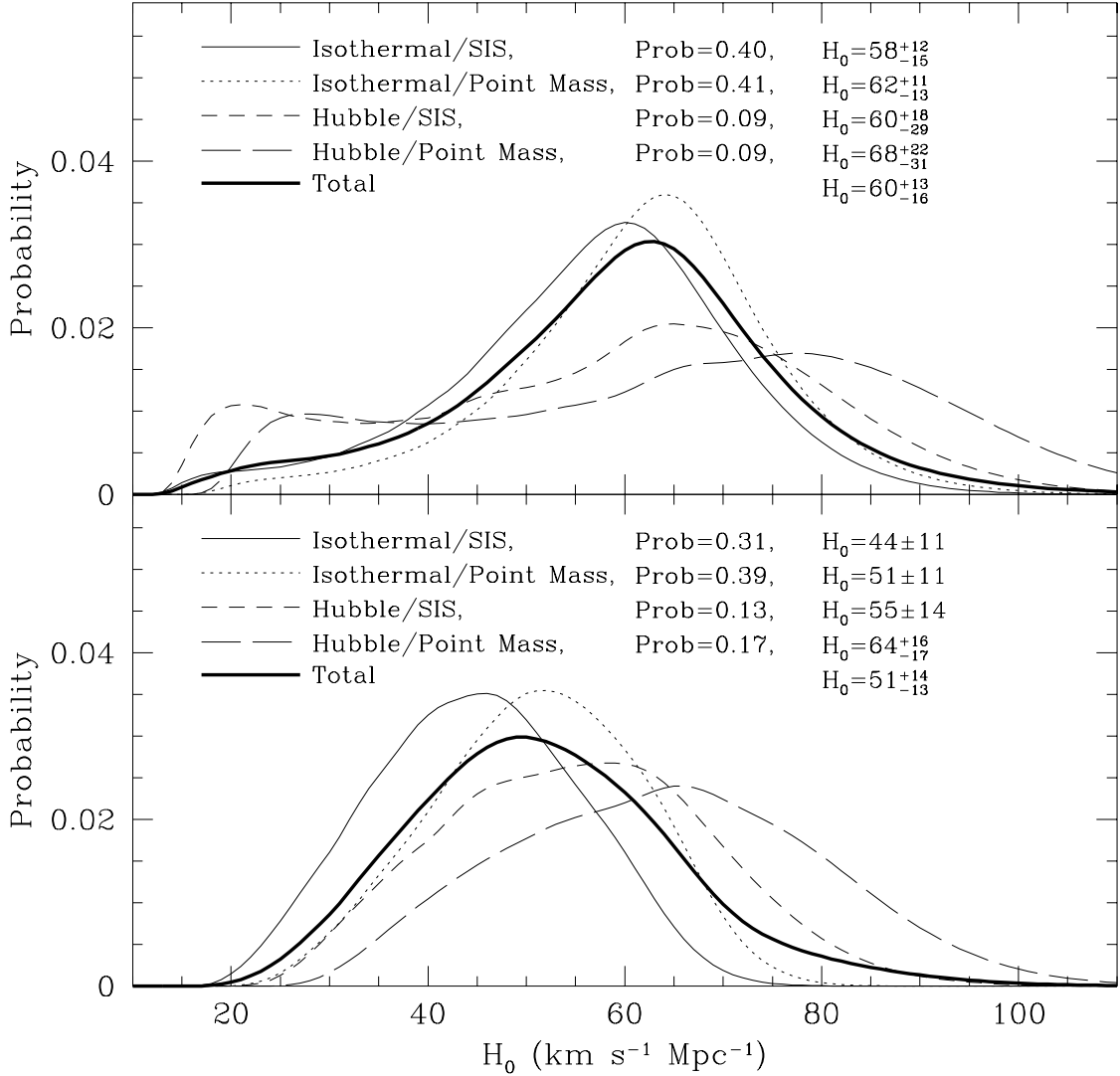}
	\caption{
Normalized probability distributions for $H_0$.  The distributions were
computed from a Bayesian analysis of the four classes of models discussed
in \S4.2.  In the top panel the Bayesian analysis does not use the constraint
from $\RABC$, and in the bottom panel it does use $\RABC$.  The relative
probabilities of the four models and their implied values for $H_0$ are
given in the key.  The spread within each distribution is due to the $6\%$
uncertainty in the observed time delay and to the degeneracy in the group
position.  The spread between the distributions is due to the degeneracies
in the profiles of the galaxy and the group.
}
	\label{fig:BayesH0}
\end{figure}

Given the wide range of group positions that provide good fits, it is
difficult to use the $\chi^2$ statistic to place limits on $H_0$.  We
can, however, use a Bayesian analysis to give a reasonable estimate of
the systematic uncertainties.  Using Bayes's theorem we can convert the
probability of the data given the parameters (given by $e^{-\chi^2/2}$)
into the probability of the parameters given the data.  We can then
compute the probability distribution for $H_0$ by integrating over the
group position, weighted by a ``prior'' probability distribution which
we took to be a circular Gaussian distribution centered on the flux
centroid C4 with standard deviation given by the rms distance of the
four galaxies from the centroid.  This is roughly the same as the
range of group positions permitted by the group mass-to-light ratio.
With the probability distribution for $H_0$ we can characterize the
systematic uncertainties in $H_0$ due to the group position, and also
estimate the relative likelihoods of the four classes of galaxy/group
profiles.  In addition, because the four classes of profiles span the
range of physically plausible models (a dark matter or constant $M/L$
galaxy, and an extended or concentrated group), we can combine them
to produce a ``total'' probability distribution that includes the
systematic effects of both the group position and the galaxy/group
profiles.  \reffig{BayesH0} shows the normalized probability
distributions and inferred values of $H_0$ for the four classes of
models, together with the total probability distribution.  Since
the time delay ratio $\RABC$ may provide a strong constraint on the
galaxy profile but is still relatively uncertain, we have computed
the probability distributions with and without the formal $\RABC$
constraint from Bar-Kana (1997).  

The Bayesian analysis emphasizes two important features of the
uncertainties in $H_0$ due to systematic uncertainties in the lens
models.  First, using two independent shears (an ellipsoidal galaxy
and a movable group) strongly affects $H_0$.  Dark matter models that
have only a single variable shear, such as the models of Schechter
et al.\ (1997) with a circular galaxy or the models of \S4.1 with the
group position fixed, give $H_0\simeq 40$ \Hunits.  By contrast, dark
matter models with two shears give $H_0=58_{-15}^{+12}$ \Hunits.  Thus
to understand the systematic uncertainties in $H_0$ it is important
to consider models with two independent shears.  Moreover, since lens
models that require two independent shears often have a degeneracy
between the shears (Keeton et al.\ 1997a), such models will generically
produce large uncertainties in $H_0$ that can be reduced only by
improving the constraints on the models.

Second, the time delay ratio $\RABC$ is an important constraint on
the profile of the galaxy.  The current estimate reduces the range of
models that are consistent with the data (especially the
models that predict large $H_0$), and reduces the probability of the
isothermal galaxy compared with the Hubble galaxy.  Without the $\RABC$
constraint, the isothermal galaxy is more likely than the Hubble galaxy
by a ratio of $4:1$, largely because the isothermal galaxy allows such
a wide range of group positions.  By contrast, with the $\RABC$ constraint
the Hubble galaxy increases in likelihood because it produces values for
$\RABC$ that are more consistent with Bar-Kana's (1997) value.  Reducing
the uncertainties in $\RABC$ thus should help discriminate between the
dark matter and constant $M/L$ galaxy models and provide the best
probe of the galaxy profile.  Using the formal $\RABC$ constraint from
Bar-Kana's (1997) present estimate, the total probability distribution
gives $H_0=51_{-13}^{+14}$ \Hunits, where these error bars incorporate
most of the systematic uncertainties in the lens models due to the group
position and the galaxy/group profiles.

\section{Stellar Dynamics}

To this point we have evaluated models purely on their ability to fit
the gravitational lens data, and we have found that the data leave
significant degeneracies in the models.  One degeneracy is related to
the galaxy profile, but here we can apply independent constraints to
try to break the degeneracy.  Specifically, we can consider whether
the galaxies required to fit the lensing data are consistent with
stellar dynamics.  For example, we suspect that the centrally
concentrated models---the models that allow high values for $H_0$---may
be unphysical.

The absence of information on the optical structure of the lens galaxy
limits the conclusions that can be drawn from stellar dynamical models
because such models require the luminosity distribution to estimate the
stellar velocities.  We assume that the galaxy is an early-type galaxy
because of its high mass and the expected dominance of lens statistics
by early-type galaxies.  Early-type galaxies are known to obey a strong
correlation between velocity dispersion, effective radius, and magnitude
known as the fundamental plane (Djorgovski \& Davis 1987; Dressler et
al.\ 1987), and we can estimate whether the lens galaxy as constrained
by the lens models can lie on the fundamental plane.  Kochanek (1993,
1996), Breimer \& Sanders (1993), and Grogin \& Narayan (1996) have
previously explored using stellar dynamical models as an added check on
lens models.

\begin{figure}[h]
	\plotone{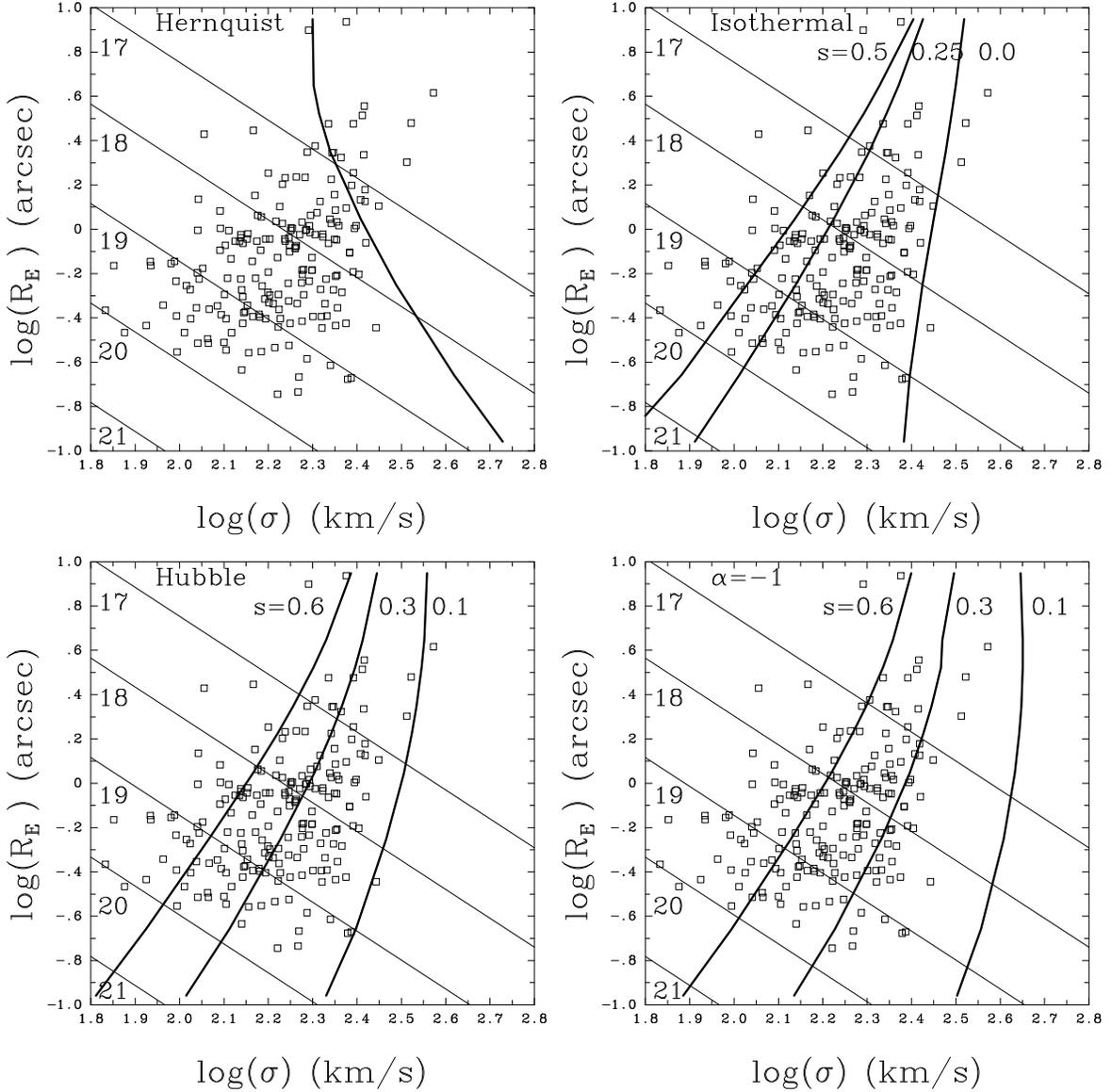}
	\caption{
Stellar dynamical limits.  The points show the effective radii and
central dispersions of the JFK sample of cluster galaxies, with the
effective radius rescaled to the redshift of \1115.  The
heavy solid lines show the central dispersion for the lens galaxy
assuming the same metric aperture as a function of $r_e$; for the
softened power-law models, the core radius $s$ is given in arcsec.
The light solid lines show the F785LP magnitude estimates for the
lens galaxy from the fundamental plane.  The K93 value
$I({\rm F785LP})=18.4$ is a central aperture magnitude and represents
a lower bound.  We found that the K93 images are consistent with a
galaxy having $r_e\simeq1\parcs5$ and $I({\rm F785LP})\simeq17$.
}
	\label{fig:dynamics}
\end{figure}

For our comparison sample we adopted the data of J\o rgensen, Franx
\& Kj\ae rgaard (1995a,b, hereafter JFK).  \reffig{dynamics} shows
the JFK sample in the space of $\log \sigma$ and $\log r_e$ shifted
to the redshift of \1115.  The velocity dispersion $\sigma$ was
estimated for a fixed metric aperture of $3\parcs4$ in diameter at
Coma (corresponding to $0\parcs4$ at \1115\ for $\Omega_0=1$).  In
order to compare the K93 and JFK photometry, we converted the JFK
Gunn $r$ magnitudes to the K93 F785LP band by using the galaxy evolution
models of Bruzual \& Charlot (1993) to compute the $K$ and evolutionary
corrections and the $r-{\rm F785LP}$ color (see Keeton et al.\ 1997b
for details).  We assumed that the luminosity density can be modeled
by a Hernquist (1990) distribution with Hernquist scale length
$a=0.45 r_e$, and we calculated only isotropic stellar dynamical
models.  From the models in \S4.2, we know that with an SIS group the
mass inside the ring of images (radius $r_0=1\parcs14=3.16h^{-1}$ kpc)
is $1.22\times10^{11} h^{-1}\ M_\odot$, with a $5\%$ uncertainty due
to the position of the group, a $10\%$ uncertainty (upwards) due to
making the group more centrally concentrated, and a $7\%$ uncertainty
(upwards) due to the cosmological model.

\reffig{dynamics} superimposes the predicted \1115\ aperture
velocity dispersion as a function of effective radius on the JFK
sample.  The self-gravitating Hernquist (constant $M/L$) models have
diverging central velocity dispersions for small effective radii
because the mass is fixed by the lens model, forcing $\sigma \propto
r_e^{-1/2}$ for effective radii smaller than the ring defined by the
images.  The effective radius is restricted to the rough range
$1\parcs0 \lesssim r_e \lesssim 4\parcs0$ if the galaxy is to lie
on the fundamental plane; then from \S4.1 the de Vaucouleurs models
with fixed group give $0.5 \lesssim \h \lesssim 0.7$ with an SIS
group and $0.6 \lesssim \h \lesssim 0.9$ with a point mass group.
Surprisingly, the dark matter models tend to have dispersion
estimates that move along the fundamental plane.  Nonetheless, many
of the centrally-concentrated lens models that lead to high values
for $H_0$ require a galaxy off the fundamental plane; for example,
the $\al=-1$ model with $s=0\parcs1$ that produces $\h=0.77$ lies
to the right of the JFK sample.  These fundamental plane constraints
can be made quantitative once we know the lens galaxy structure and
effective radius.

Note that the Hernquist models require a total lens galaxy magnitude
of $I({\rm F785LP})\simeq17$ in order to lie on the fundamental plane,
whereas K93 estimated a central aperture magnitude for the lens
galaxy of $I({\rm F785LP})=18.36$ mag.  We experimented with the
K93 images and found that, due to the wings of the original WFPC
point-spread function and to the huge contrast between the quasar
and galaxy surface brightnesses, the data are consistent with lens
models having $I({\rm F785LP})\simeq17$ and
$r_e\simeq1\parcs5$--$2\parcs0$.  In addition, Keeton et al.\ (1997b)
have remarked that the K93 magnitude estimate gives a luminosity well
below that expected from the ``Faber--Jackson'' type relation between
image separation and lens luminosity that other lenses obey.  A higher
luminosity for the \1115\ lens galaxy would move it closer to the
trend followed by other lenses.  Improved HST photometry will reveal
whether this is indeed the case.

Although this analysis is limited at present by the lack of optical
data on the lens galaxy, it does reveal that the centrally-concentrated
galaxy models that give high values for $H_0$ are physically implausible.
The dark matter models are consistent with stellar dynamics, as are the
constant $M/L$ models if the effective radius is in the range
$1\parcs0 \lesssim r_e \lesssim 4\parcs0$.

\section{Discussion}

By combining a lens model of \1115\ with the observed time delays of
Bar-Kana (1997), we can infer a value for $H_0$ that is independent
of the standard distance ladder.  The resulting value depends on the
lens model, so we explored a range of models to understand the systematic
uncertainties in the models, their effects on $H_0$, and the types of
future observations that can reduce the uncertainties.  We found that
\1115\ cannot be fit by an isolated lens galaxy, but that it can be
well fit ($\chiN<1$) by including the effects of the nearby group.
To understand the full range of model uncertainties, it is important
to use an ellipsoidal galaxy and a movable group.  Since the present
observational data do not constrain the galaxy and group profiles or
the group position, we studied various profiles and group positions
and found good fits with $H_0$ ranging from roughly $30$ to $90$
\Hunits.  However, many of these models are unacceptable on physical
grounds.  For example, some of the models that give high values for
$H_0$ require a galaxy mass profile that is more centrally concentrated
than typical luminosity profiles and correspond to a galaxy that does
not lie on the fundamental plane.  If we restrict the models to a
plausible constant $M/L$ model and a dark matter model, we can use a
Bayesian analysis to characterize the systematic uncertainties in
$H_0$ relating to degeneracies in the models.  We find
$H_0=51_{-13}^{+14}$ \Hunits, where the error bars incorporate the
uncertainty in the measured time delays as well as the uncertainties
in the lens models due to the group position and the galaxy and group
profiles.  They do not, however, take into account the systematic
uncertainties in the value of the time delay ratio $\RABC$ from
Bar-Kana (1997).  In addition to the formal uncertainties, there may
be a $5$--$10\%$ uncertainty due to mass fluctuations from large-scale
structure (Bar-Kana 1996; Wambsganss et al.\ 1997), and a $7\%$
uncertainty (upwards) due to the cosmological model (\reftab{cosmology}).

The uncertainties in $H_0$ can be reduced with better observational
constraints, but perhaps surprisingly not with better relative quasar
positions and fluxes.  In most models the quasar positions are already
overfit, and an order of magnitude improvement in the uncertainties
(to $0.5$ mas) is not practical.  The quasar fluxes, by contrast, are
dominated by systematic effects from microlensing and extinction rather
than by measurement errors.  We accounted for these effects by using
flux error bars of $20\%$; merely reducing the flux error bars by a
factor of two changes the absolute $\chi^2$ but has little effect on
the best-fit model parameters.  Improving the constraints from the
quasar fluxes would require understanding the systematic effects
of microlensing and extinction, and in particular understanding why
the K93 value of the $A_2/A_1$ flux ratio ($0.66$ in $V$ and $0.70$
in $I$) differs from the best-fit models (typically $\gtrsim0.90$)
and from theoretical expectations of a value near unity.  One way to
avoid problems with microlensing and extinction would be to measure
radio flux ratios; unfortunately \1115\ is not a strong radio
source (flux $<1.5$ mJy; Weymann et al.\ 1980).

The most promise for improving the constraints comes from the galaxy
position and the time delays (see \S4.2).  Reducing the error bars on
the galaxy position from $50$ mas to $10$ mas or better will rule out
many models, and such a reduction should be possible with new WFPC2
observations.  Reducing the error bars on the time delays by a factor
of four (to $\sim0.5$ day in the time delays or to $\sim 0.05$ in
$\RABC$) will provide a useful independent constraint.  In fact, the
time delay ratio $\RABC$ may provide the best probe of the galaxy
profile.  The effects of the galaxy position and the time delay ratio
can be seen by noting that, together with the convergence at the ring
of images, they account for almost all of the variation in $H_0$ from
model to model.  If we combine the models from \S4.1 (different galaxy
profiles with varying core radius and fixed group) and the models from
\S4.2 (singular isothermal ellipsoid or modified Hubble model galaxy,
with movable group), then the implied Hubble constant is correlated
with the galaxy position and the time delay ratio,
\begin{equation}
	\h = \biggl[0.83 + 0.005 \Delta x_{gal} - 0.006 \Delta y_{gal} +
		0.10 (\RABC-1.4) \biggr] (1-\kappa_0)
\end{equation}
where $\Delta x_{gal}$ and $\Delta y_{gal}$ are the galaxy position in
mas relative to that of K93, and $\kappa_0=\kgal+\kgrp$ is the total
convergence at the ring of images due to the galaxy and the group.  For
a singular isothermal galaxy $\kgal=1/2$ at the ring of the images; for
an SIS group $\kgrp=\ggrp\sim10\%$ in \1115, while for a point mass
group $\kgrp=0$.  This correlation is meant only as a qualitative guide
to the quantities that will best constrain the Hubble constant, and it
does not address the degree to which the convergence $\kappa_0$ can be
determined from improved data.

If even with improved constraints we still cannot determine the galaxy
and group profiles and the group position directly, then perhaps we
can impose external considerations.  First, as in \S5 we can use
stellar dynamics and the fundamental plane to identify which models are
physically plausible.  At present this analysis can only rule out very
concentrated models, but with luminosity profile data for the galaxy it
may provide a more quantitative constraint.  Unfortunately, directly
measuring the central velocity dispersion in \1115\ will be very
difficult due to the high quasar/galaxy contrast.  Second, we could
require consistency with other gravitational lens statistics and
models (Grogin \& Narayan 1996; Kassiola \& Kovner 1993; Kochanek 1993,
1995, 1996, Wallington \& Narayan 1993; Maoz \& Rix 1993) and with
observations of elliptical galaxies (Fabbiano 1989; Rix 1996) to say
that lens galaxies have significant dark matter and small core radii.
In this case, the models from \S4.2 give $H_0=44\pm11$ \Hunits\ using
the $\RABC$ constraint.  Finally, there are now two lenses (0957$+$561
and \1115) for which time delays have been measured.  Both systems
exhibit a degeneracy between the mass distribution and the Hubble
constant (e.g.\ Grogin \& Narayan 1996), but by requiring the lenses
to agree on both the Hubble constant and the typical mass distributions
of galaxies we may be able to break the degeneracies.

\acknowledgments

Acknowledgments:  We thank P.~Schechter, R.~Bar-Kana, E.~Falco,
J.~Leh\'ar, R.~Narayan, B.~McLeod, M.~Franx, and M.~Geller for
useful discussions.  CRK is supported by ONR-NDSEG grant
N00014-93-I-0774.  CSK is supported by NSF grant AST-9401722.

\clearpage

\begin{deluxetable}{clcccc}
\tablecaption{Cosmological scale factors}
\tablehead{
 & \multicolumn{1}{r}{$(\Omega_0,\lambda_0)=$} & $(1,0)$ & $(0.1,0)$ & $(0.4,0.6)$ & $(0.2,0.8)$
}
\startdata
$1\arcs$ & ($h^{-1}$ kpc)                                               & 2.771 & 2.968 & 3.080 & 3.219 \\
${D_l D_s\over 2r_H D_{ls}}$ &                                          & 0.139 & 0.149 & 0.145 & 0.145 \\
${(1+z_l)\over c}{D_l D_s\over D_{ls}}$ & ($h^{-1}$ days arcsec$^{-2}$) & 30.49 & 32.73 & 31.68 & 31.80 \\
$\Sigma_{cr}$ & ($10^{11} h^{-1}\,M_\odot\ {\rm arcsec}^{-2}$)          & 0.326 & 0.350 & 0.339 & 0.340 \\
$\Sigma_{cr}$ & ($h$ g cm$^{-2}$)                                       & 0.888 & 0.831 & 0.747 & 0.686 \\
\enddata
\tablecomments{Quantities are computed for $z_s=1.722$, $z_l=0.304$.  The first
two cosmological models are FRW cosmologies with the specified $\Omega_0$.
The last two cosmological models are flat cosmologies with $\Omega_0+\lambda_0=1$.}
\label{tab:cosmology}
\end{deluxetable}

\begin{deluxetable}{lcccc}
\tablecaption{Observational data}
\tablehead{
   & $x$ (\arcs) & $y$ (\arcs) & F555W (mag) & F785LP (mag)
}
\startdata
A1 & $  -1.294$  & $  -2.036$  & $16.90$     & $16.12$ \\
A2 & $  -1.448$  & $  -1.582$  & $17.35$     & $16.51$ \\
B  & $\s 0.362$  & $  -1.949$  & $18.87$     & $18.08$ \\
C  & $\s 0.000$  & $\s 0.000$  & $18.37$     & $17.58$ \\
G  & $  -0.355$  & $  -1.322$  & $     $     & $18.36$ \\
\enddata
\tablecomments{Data from HST WFPC observations by Kristian et al.\ (1993).
$x$ is approximately west, $y$ is approximately north.  The internal
position uncertainties are $5$ mas for the quasar images and $50$ mas
for the galaxy.  Formally, the relative fluxes are uncertain by $1.5\%$
in $I$ (F785LP) and $3\%$ in $V$ (F555W), while the zero-point for
magnitudes is uncertain by $0.3$ mag.
The positions and Gunn $r$ magnitudes of the group galaxies are as
follows (Young et al.\ 1981):
G1$=(\dgrp=23\parcs5, \tgrp=-118\deg, r=18.96)$,
G2$=(12\parcs0,  -95\deg, 20.04)$,
G3$=(18\parcs9, -131\deg, 20.53)$.
}
\label{tab:data}
\end{deluxetable}

\begin{deluxetable}{rccccc}
\tablecaption{Galaxy + fixed group models}
\tablehead{& \multicolumn{1}{r}{Galaxy:} & de Vaucouleurs & Isothermal & Modified Hubble & $\al=-1$ }
\startdata
SIS Group  & $r_e$ or $s$ (\arcs) & $1.00$ & $0.00$ & $0.19$ & $0.42$ \\
           & $\h$                 & $0.65$ & $0.41$ & $0.65$ & $0.75$ \\
           & $\RABC$              & $1.40$ & $1.35$ & $1.40$ & $1.44$ \\
           & $b/a$                & $0.87$ & $0.97$ & $0.87$ & $0.82$ \\
           & $\Delta_{gal}$ (mas) & $27.5$ & $12.3$ & $29.5$ & $44.4$ \\
           & $\chi^2_{pos}$       & $0.04$ & $0.01$ & $0.06$ & $0.53$ \\
           & $\chi^2_{flux}$      & $1.64$ & $2.27$ & $1.60$ & $1.49$ \\
           & $\chi^2$             & $1.98$ & $2.34$ & $2.01$ & $2.82$ \\
\tableline
Point Mass & $r_e$ or $s$ (\arcs) & $3.00$ & $0.00$ & $0.27$ & $0.47$ \\
Group      & $\h$                 & $0.68$ & $0.51$ & $0.83$ & $0.97$ \\
           & $\RABC$              & $1.40$ & $1.36$ & $1.43$ & $1.47$ \\
           & $b/a$                & $0.84$ & $0.90$ & $0.78$ & $0.73$ \\
           & $\Delta_{gal}$ (mas) & $41.4$ & $27.2$ & $55.3$ & $69.3$ \\
           & $\chi^2_{pos}$       & $0.07$ & $0.03$ & $0.35$ & $2.19$ \\
           & $\chi^2_{flux}$      & $1.98$ & $2.31$ & $1.77$ & $1.98$ \\
           & $\chi^2$             & $2.74$ & $2.63$ & $3.34$ & $6.09$ \\
\enddata
\tablecomments{Results from the best-fit galaxy + fixed group models.
The group position is fixed at the flux-weighted centroid of the four
galaxies (C4 in Figure 1).  The de Vaucouleurs models were computed
only for discrete $r_e$ spaced by $0\parcs5$.  The softened power-law
models were computed with $s$ varying continuously.  The $\chi^2_{pos}$
and $\chi^2_{flux}$ indicates the contribution to the $\chi^2$ from the
quasar positions and fluxes, respectively.  Each model with fixed $r_e$
or $s$ has $N_{dof}=5$.  We do not give error bars because they would
not encompass the effects of allowing the group to move.
}
\label{tab:results}
\end{deluxetable}

\clearpage

\end{document}